\documentclass[%
APS,
rsi,%
 amsmath,amssymb,superscriptaddress,
reprint,
]{revtex4-2}
\usepackage{lipsum}
\usepackage{subfigure}
\usepackage{graphicx}
\usepackage{dcolumn}
\usepackage{placeins}
\usepackage{bm}
\usepackage{siunitx}
\usepackage{gensymb}
\usepackage{xcolor}
\usepackage{floatpag}
\usepackage{float}

\usepackage{graphicx}
\usepackage[export]{adjustbox}
\allowdisplaybreaks

\graphicspath{ {figures/} }
\begin{document}

\preprint{AIP/123-QED}

\title[Sample title]{Magnon-magnon coupling in synthetic ferrimagnets}
\author{A. Sud}
 \email{aakanksha.sud.c1@tohoku.ac.jp}
 \affiliation{WPI Advanced Institute for Materials Research, Tohoku University, 2-1-1, Katahira, Sendai 980-8577, Japan}
\author{K. Yamamoto}
 \affiliation{Advanced Science Research Center, Japan Atomic Energy Agency, 2-4 Shirakata, Tokai 319-1195, Japan}
 \author{K. Z. Suzuki}
 \affiliation{WPI Advanced Institute for Materials Research, Tohoku University, 2-1-1, Katahira, Sendai 980-8577, Japan}
  \affiliation{Advanced Science Research Center, Japan Atomic Energy Agency, 2-4 Shirakata, Tokai 319-1195, Japan}
\author{S. Mizukami}
 \affiliation{WPI Advanced Institute for Materials Research, Tohoku University, 2-1-1, Katahira, Sendai 980-8577, Japan}
 \affiliation{Center for Science and Innovation in Spintronics, Tohoku University, Sendai, 980-8577, Japan}
\author{H. Kurebayashi}%
   \affiliation{WPI Advanced Institute for Materials Research, Tohoku University, 2-1-1, Katahira, Sendai 980-8577, Japan}
\affiliation{London Centre for Nanotechnology, University College London, London WC1H 0AH, United Kingdom}
\affiliation{Department of Electronic and Electrical Engineering, University College London, London, WC1E 7JE, United Kingdom}
\date{\today}

\begin{abstract}
Magnetic multilayers with interlayer exchange coupling have been widely studied for both static and dynamic regimes. Their dynamical responses depend on the exchange coupling strength and magnetic properties of individual layers. Magnetic resonance spectra in such systems are conveniently discussed in terms of coupling of acoustic and optical modes. At a certain value of applied magnetic field, the two modes come close to being degenerate and the spectral gap indicates the strength of mode hybridisation. In this work, we theoretically and experimentally study the mode hybridisation of interlayer-exchange-coupled moments with dissimilar magnetisation and thickness of two ferromagnetic layers. In agreement with symmetry analysis for eigenmodes, our low-symmetry multilayers exhibit sizable spectral gaps for all experimental conditions. The spectra agree well with the predictions from the Landau-Lifshitz-Gilbert equation at the macrospin limit whose parameters are independently fixed by static measurements.  
\end{abstract}
 \maketitle
 \section{\label{sec:level1}Introduction\protect\\}
 
In two magnetic layers separated by a thin nonmagnetic spacer, conduction electrons in the spacer magnetically couple two spatially separated moments, via the so-called Ruderman-Kittel-Kasuya-Yosida (RKKY) interaction~\cite{Ruderman_PR1954,Kasuya_PTP1956,Yoshida_PR1957}. This interlayer exchange coupling arises from coherent propagation of electron spin across the spacer layer~\cite{Bruno_PRB1995,Slonczewski_PRB1989,Edwards_PRL1991}. Due to the Friedel-like oscillation of the electron phase, the exchange coupling changes its sign as a function of the interlayer distance, switching between ferromagnetic and antiferromagnetic ordering of the two magnetic layers~\cite{Grunberg_PRL1986,Majkrzak_PRL1986,Salamon_PRL1986,Parkin_PRL1990}. The antiferromagnetically ordered states of two identical magnetic layers, often called synthetic antiferromagnets (SyAFs), have served as a testbed for studying antiferomagnetism where SyAFs' relatively weak RKKY exchange coupling, comparable to the strength of magnetic fields achievable in laboratories, helps realise experiments otherwise difficult in atomically-ordered, crystalline antiferromagnets~\cite{Jungwirth_NatNano2016,Duine_NatPhys2018,Baltz_RMP2018}. One such property is the magnetic resonances in SyAFs whose typical frequency resides within a range of GHz that is readily accessible by modern microwave techniques~\cite{PhysRevB.21.169,Zhang_PRL1994,Zhang_PRB1994,Belmeguenai_PRB2007,Topkaya_JAP2010,doi:10.1063/1.3143625,Khodadadi_PhysRevAppli2017,PhysRevB.90.064428,Sorokin_PRB2020,sud2020tunable,He_CPL2021,Patchett_PRB2022,Waring_PRB2021,Wang_CPB2022}. 

In a canted static state in SyAFs under an applied magnetic field, the low-lying, spatially-uniform resonance modes are usually called acoustic and optical modes where the precessions of two exchange-coupled moments are primarily in- and out-of-phase, respectively~\cite{rezende1998studies}. The resonance frequencies of these modes exhibit different magnetic field dependence, allowing them to come almost degenerate in a certain field range. Unless some symmetry conditions are satisfied, there is no exact degeneracy and the spectrum is gapped as a function of magnetic field~\cite{shiota2020tunable,sud2020tunable}. In the regime where the two modes have well-defined in- and out-of-phase characters away from the degenerate range, the gap, defined as the minimum split with respect to the field between the two resonance frequencies, represents the coupling strength of the in- and out-of-phase oscillations and when it is greater than the linewidths, it means that the energy transfer between them takes place more frequently before the excited state is lost, termed as the strong coupling regime~\cite{ZARERAMESHTI_PhysRep2022}. A large coupling strength has been favoured for potential magnonic applications in which such energy transfer might play a crucial role~\cite{YUAN_PhysRep2022,ZARERAMESHTI_PhysRep2022,Awschalom_IEEETQM2021,Chumak_IEEE2022}. 

Coupled-moment systems offer unique research directions and potential spintronic applications~\cite{Jungwirth_NNano2016,duine2018synthetic,Kim_NMater2022,Han_NMater2023,Wang_APLMater2023}. For example, their tunable material parameters enable the control of characteristic frequencies in nano-oscillator devices~\cite{Houssameddine_JPhysD2010,Volvach_PRMater2022}, up to the THz regime~\cite{Zhong_JMMM2020}. The oscillation of magnetisations in SyAFs could be excited by spin-orbit torques (SOTs) in planar geometories~\cite{Lau_NNano2016,sud2021parity}. SOTs in turn drive fast domain-wall propagation/dynamics in compensated magnets~\cite{Yang_NNano2015}. Due to the compensated nature, skyrmions in such a system~\cite{Legrand_NMater2020} benefit from the cancellation of the skyrmion Hall effect, moving straight within a propagation channel~\cite{Zhang_NComm2016,Dohi_NComm2019}. We also envisage that some of unique properties in coupled-moment systems can be used in exploring a variety of neuromorphic computation schemes~\cite{Grollier_NElec2020,Allwood_APL2023,Lee_APL2023}.

\begin{figure*}[ht!]
\centering
\includegraphics[width=17cm]{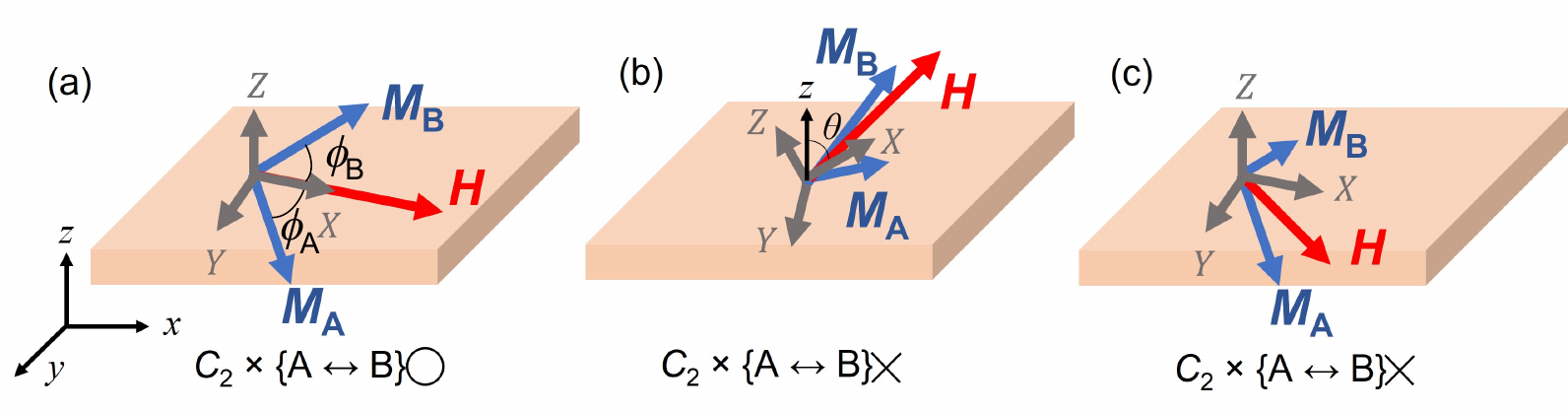}
\caption{(a) In the laboratory frame, we define the $z$ direction normal to the plane, and $x$ direction such that the static external magnetic field lies in the $x$-$z$ plane. In the canted regime when applying the field ($\bm{H}$) in-plane, two sub-lattice moments ($\bm{M}_A$ and $\bm{M}_B$) reside within the plane, canted towards $\bm{H}$. For general static states, we introduce new coordinate axes $\bm{X},\bm{Y},\bm{Z}$ adapted to the two-fold rotation $\mathcal{C}_2$ that brings the unit vector along $\bm{M}_A$ to that along $\bm{H}$. See Eq.~(\ref{eq:frame}) for the concrete definition. For $\bm{H}$ in-plane and identical magnetic layers, $\mathcal{C}_2$ combined with interchanging A and B layers is a symmetry of the system. (b) and (c) When we apply $\bm{H}$ with the polar angle $\theta\neq 90\degree$ or two magnetic moments are not identical, $\mathcal{C}_2$ followed by the magnetic layer interchange ceases to be a symmetry. This impacts on the mode coupling as discussed in this study.}
\label{fig1}
\end{figure*}

In the most commonly studied setup of magnetisation dynamics in SyAFs, however, the in- and out-of-phase oscillations are seen to remain eigenmodes for any magnetic field value and become fully degenerate, where symmetry of the system plays a crucial role for the decoupling~\cite{macneill2019gigahertz,PhysRevB.103.064429,Patchett_PRB2022,Dai_APL2021}. MacNeill \textit{et al.} present that the coupled Landau-Lifshitz-Gilbert (LLG) equations due to the interlayer exchange interaction are symmetric under twofold rotation around the applied field direction combined with a layer swap, as long as the field is within the film plane (Fig.~\ref{fig1}(a))~\cite{macneill2019gigahertz}. The acoustic and optical modes are odd and even upon the symmetry operation respectively and therefore unable to hybridise with each other, leading to the mode degeneracy at a resonance point~\cite{PhysRevB.21.169,rezende1998studies}. This specific symmetry can be broken in several ways, for example, tilting the external magnetic field towards out-of-plane direction (Fig.~\ref{fig1}(b))~\cite{macneill2019gigahertz, sud2020tunable, PhysRevB.103.064429}. For general expressions of spin-wave mode frequencies in interlayer exchange-coupled systems, Layadi presented analytical solutions with a particular focus on the effect of the biquadratic exchange coupling and in-plane anisotropy on the spectra for in-plane magnetised cases~\cite{Layadi_PRB20002}. While spectroscopic measurements of interlayer exchange-coupled tri-layers with different magnetic-layer thicknesses were reported by some groups in the past~\cite{Belmeguenai_JPCM2008,Topkaya_JAP2010,liu2014ferromagnetic,Sorokin_PRB2020,Gladii_PRB2023}, there seem no study fully dedicated to quantitative discussions of the mode hybridisation in such asymmetric interlayer-exchange-coupled systems.

In this paper, we present our detailed experimental and theoretical study of the magnon-magnon coupling phenomena in synthetic ferrimagnets where two magnetically coupled layers are not identical (Fig.~\ref{fig1}(c)). We systematically compare spin-wave spectra measured by broadband ferromagnetic resonance (FMR) experiments and calculated using magnetic parameters deduced from static magnetometry. In all cases examined, we find excellent agreement between experiment and theory, suggesting that the coupled LLG equations at the macrospin limit are indeed a reliable tool for designing and analysing the spectral properties of the magnetic multilayers. Our calculations further reveal dissimilar roles of quadratic and biquadratic exchange interactions for the size of the gap. Our results 
help design and control magnetic resonance spectra in exchange-coupled magnetic moments that can be synthetic antiferro(ferri)magnets, van der Waals antiferromagnets~\cite{macneill2019gigahertz,SklenarPRMater2021,Cenker_NPhys2021,Tang2023_vdWreview} and ferromagnetic bi-layers~\cite{Klingler_PRL2018,Chen_PRL2018,Li_PRL2020}.

\section{A macrospin model of synthetic ferrimagnet} 

For our purposes, a theoretical model that extends the result of Ref.~\cite{Layadi_PRB20002} for arbitrary direction of the static magnetic field is required, which we present in this section with an emphasis on breaking of the two-fold rotation symmetry. Let $\bm{M}_A , \bm{M}_B$ be the magnetisations of the two ferromagnetic layers. We are interested in the situations where the two magnetic materials are not identical $\left| \bm{M}_A \right| \equiv M_A \neq \left| \bm{M}_B \right| \equiv M_B$, and the two layers have different thicknesses $d_A \neq d_B$. The film normal is chosen $z$ axis and the film is regarded infinitely extended in the $x,y$ directions, as shown in Fig.~\ref{fig1}(a). 

The static state of the magentisations corresponds to the minimum of free energy per unit area $W$. We include the external magnetic field $\bm{H}$, demagnetising field, and biquadratic as well as the usual quadratic interlayer exchange interactions: 
\begin{align}
W =&\ d_A \left\{ -\mu _0 M_A \bm{H}\cdot \bm{n}_A + \frac{\mu _0 M_A^2}{2} \left( n_A^z \right) ^2 \right\}  \notag \\
& + d_B \left\{ -\mu _0 M_B \bm{H}\cdot \bm{n}_B + \frac{\mu _0 M_B^2}{2} \left( n^z_B \right) ^2 \right\}  \notag \\
&+  J_1 \bm{n}_A \cdot \bm{n}_B + J_2 \left( \bm{n}_A \cdot \bm{n}_B \right) ^2 . 
\label{eq:free-energy}
\end{align}
Here we have normalised the magnetisations $\bm{n}_{A(B)} = \bm{M}_{A(B)} /M_{A(B)}$, and introduced the phenomenological exchange energies per unit area $J_1$ and $J_2$. Without loss of generality, with the weak crystalline anisotropy being ignored, the magnetic field can be taken $\bm{H} = H \left( \hat{\bm{x}}\sin \theta + \hat{\bm{z}}\cos \theta \right) $. We determine the static state $\bm{n}_{A(B)}^0$ by numerical minimisation of $W$, which is parameterised by
\begin{equation}
\bm{n}^0_{A(B)} = \begin{pmatrix}
	\sin \theta _{A(B)} \cos \phi _{A(B)} \\
	\sin \theta _{A(B)} \sin \phi _{A(B)} \\
	\cos \theta _{A(B)} \\
	\end{pmatrix} .
\end{equation}
If the magnetic field is in-plane $\theta =90\degree$ and $0<2J_2 <J_1$, the static state undergoes two phase transitions at $H_{\rm sf}$ and $H_{\rm ff}$ as $\left| H \right| $ is increased from zero, where
\begin{align}
H_{\rm sf} =&\left| \frac{1}{d_B M_B}-\frac{1}{d_A M_A} \right| \frac{J_1 -2J_2}{\mu _0} , \\
H_{\rm ff} =&\left| \frac{1}{d_B M_B}+\frac{1}{d_A M_A} \right| \frac{J_1 +2J_2}{\mu _0} .
\end{align}
Below $H_{\rm sf}$, the static state is antiferromagentic $\bm{n}^0_B =-\bm{n}^0_A $ with $\bm{n}^0_A \cdot \bm{H} \gtrless 0$ according to $d_A M_A \gtrless d_B M_B$. Above $H_{\rm ff}$, the system is in a forced ferromagnetic state $\bm{n}^0_A =\bm{n}^0_B = \bm{H}/\left| H\right| $. In between lies the spin-flop, or canted, state where $H\cos \phi _{A,B}> 0, \sin \phi _A \sin \phi _B <0$. 

To calculate the magnetic resonance frequencies, let us introduce the linear perturbation $\bm{n}_{A(B)} \approx \bm{n}^0_{A(B)} + \bm{n}^1_{A(B)} $ where $\bm{n}^0_{A(B)} \cdot \bm{n}^1_{A(B)} =0$. The Landau-Lifshitz equations follow from the free energy $W$ through the usual procedure~\cite{StancilBook}. Although one can press on using $\bm{n}_{A(B)}^1$ as the dynamical variables~\cite{PhysRevB.103.064429}, we normalise them so as to make them canonical in the sense of Hamiltonian mechanics~\cite{StancilBook}, which ensures that the resulting eigenvalue problem retains the correct Bogoliubov form~\cite{Colpa1978}:
\begin{equation}
\bm{\delta }_A = \sqrt{\frac{Sd_A M_A}{\hbar \left| \gamma _A \right|}}\bm{n}_A^1 , \quad \bm{\delta }_B =\sqrt{\frac{Sd_B M_B}{\hbar \left| \gamma _B \right|}} \bm{n}_B^1 , \label{eq:normalisation}
\end{equation}
where $S$ denotes the area of the film, and $\gamma _{A(B)}<0$ are the gyromagnetic ratios. The linearised equations of motion read
\begin{widetext}
\begin{align}
\bm{n}_A^0 \times \frac{d\bm{\delta }_A}{dt} =&\ \gamma _A \mu _0 \left[ \left\{ \bm{H}\cdot \bm{n}_A^0 - M_A \left( \hat{\bm{z}}\cdot \bm{n}_A^0 \right) ^2  \right\} \bm{\delta }_A + M_A \left( \hat{\bm{z}}\cdot \bm{\delta }_A \right) \left\{ \hat{\bm{z}} - \left( \hat{\bm{z}}\cdot \bm{n}_A^0 \right) \bm{n}_A^0 \right\}  \right] \nonumber \\
& -\frac{\gamma _A}{d_A M_A} \left\{ J_1 + 2\left( \bm{n}_A^0 \cdot \bm{n}_B^0 \right) J_2 \right\} \left[  \left( \bm{n}_A^0 \cdot \bm{n}_B^0 \right) \bm{\delta }_A - \sqrt{\frac{\gamma _B dA M_A}{\gamma _A d_B M_B}} \left\{ \bm{\delta }_B - \left( \bm{n}_A^0 \cdot \bm{\delta }_B \right) \bm{n}_A^0  \right\} \right] \notag \\
& +\frac{2\gamma _A }{d_A M_A} J_2 \left\{ \bm{n}_B^0 - \left( \bm{n}_A^0 \cdot \bm{n}_B^0 \right) \bm{n}_A^0 \right\} \left( \bm{n}_B^0 \cdot \bm{\delta }_A + \sqrt{\frac{\gamma _B d_A M_A}{\gamma _A d_B M_B}} \bm{n}_A^0 \cdot \bm{\delta }_B \right) , \label{eq:nA}  \\
\bm{n}_B^0 \times \frac{d\bm{\delta }_B}{dt} =&\ \gamma _B \mu _0 \left[  \left\{ \bm{H}\cdot \bm{n}_B^0 - M_B \left( \hat{\bm{z}}\cdot \bm{n}_B^0 \right) ^2 \right\} \bm{\delta }_B + M_B \left( \hat{\bm{z}}\cdot \bm{\delta }_B \right) \left\{ \hat{\bm{z}} -\left( \hat{\bm{z}}\cdot \bm{n}_B^0 \right) \bm{n}_B^0 \right\} \right] \notag \\
& - \frac{\gamma _B}{d_B M_B}    \left\{ J_1 + 2\left( \bm{n}_A^0 \cdot \bm{n}_B^0 \right) J_2 \right\} \left[ \left( \bm{n}_A^0 \cdot \bm{n}^0_B \right) \bm{\delta }_B - \sqrt{\frac{\gamma _A d_B M_B}{\gamma _B d_A M_A}} \left\{ \bm{\delta }_A - \left( \bm{n}_B^0 \cdot \bm{\delta }_A \right) \bm{n}_B^0 \right\} \right] \notag \\
& + \frac{ 2 \gamma _B }{d_B M_B} J_2  \left\{ \bm{n}_A^0 - \left( \bm{n}_A^0 \cdot \bm{n}_B^0 \right) \bm{n}_B^0 \right\} \left( \bm{n}_A^0 \cdot \bm{\delta }_B + \sqrt{\frac{\gamma _A d_B M_B}{\gamma _B d_A M_A}} \bm{n}_B^0 \cdot \bm{\delta }_A \right). \label{eq:nB} 
\end{align}
\end{widetext} 

Equations (\ref{eq:nA}) and (\ref{eq:nB}) describe two coupled harmonic oscillators, i.e. there are four independent real functions of time to be determined. We are interested in the resonance properties, which can be analyzed in terms of any consistent choice of the four independent variables. Had it not been for the shape anisotropy and the asymmetry between $d_A, M_A ,\gamma _A $ and $d_B , M_B ,\gamma _B$, two-fold rotation around $\bm{H}$ would have mapped $\bm{n}_A^0 $ to $\bm{n}_B^0 $ and the symmetry-adapted variables would have been convenient. Following MacNeil \emph{et al.}~\cite{macneill2019gigahertz}, let $\mathcal{C}_2$ denote the two-fold rotation that brings $\bm{n}_A^0$ to $\bm{n}_B^0 $ whose axis coincides with $X$ direction in Fig.~\ref{fig1}. Algebraically the action of $\mathcal{C}_2$ on an arbitrary vector $\bm{v}$ is given by
\begin{equation}
\mathcal{C}_2 \bm{v} = \frac{\left( \bm{n}_A^0 + \bm{n}_B^0 \right) \cdot \bm{v}}{1+\bm{n}_A^0 \cdot \bm{n}_B^0 }\left( \bm{n}_A^0 +\bm{n}_B^0 \right) -\bm{v} . \label{eq:rotation}
\end{equation}
Although $\mathcal{C}_2$ is not in general a symmetry of the problem, it helps make sense of the results in terms of the familiar notions used in previous studies~\cite{macneill2019gigahertz,sud2020tunable}. The definition of $\mathcal{C}_2$ becomes ambiguous for $\left| H \right| < H_{\rm sf} $ and what follows does not work for $\left| H\right| >H_{\rm ff}$ either, but these collinear cases are simple and separately handled in the Appendix. Focusing on the spin-flop phase, we introduce $\bm{\delta }_{\pm } = \left( \bm{\delta }_A \pm \mathcal{C}_2 \bm{\delta }_B \right) /\sqrt{2}$ that are even and odd eigenvectors of $\mathcal{C}_2 \times \{ A\leftrightarrow B \}$. To pick out two independent components each for $\bm{\delta }_{\pm }$, we define a new coordinate frame $XYZ$ (Fig.~\ref{fig1}) given by
\begin{equation}
\hat{\bm{X}} = \frac{\bm{n}_A^0 + \bm{n}_B^0}{\sqrt{2+2\bm{n}_A^0 \cdot \bm{n}_B^0}} , \quad \hat{\bm{Y}} = \frac{\bm{n}_A^0 -\bm{n}_B^0}{\sqrt{2-2\bm{n}_A^0 \cdot \bm{n}_B^0}} , \label{eq:frame}
\end{equation}
and $\hat{\bm{Z}} =\hat{\bm{X}}\times \hat{\bm{Y}}$. By construction, $\bm{n}_A^0 \cdot \bm{\delta }_{\pm }=0$ so that one may write $\bm{\delta }_{\pm } =\delta _{\pm }^{\perp Z } \hat{\bm{Z}}\times \bm{n}_A^0 + \delta _{\pm }^{\parallel Z }\hat{\bm{Z}}$. As is usual in cavity magnonics, we work with the complex variables $\alpha = \delta _-^{\perp Z } -i\delta _-^{\parallel Z } , \beta = \delta _+^{\perp Z } -i\delta _+^{\parallel Z }$ that would represent annihilation operators in the quantum regime. This change of variables brings Eqs.~(\ref{eq:nA}) and (\ref{eq:nB}) into
\begin{widetext}
\begin{equation}
i\frac{d}{dt} \begin{pmatrix}
	\alpha \\
	-\overline{\alpha } \\
	\beta \\
	-\overline{\beta } \\
	\end{pmatrix} = \begin{pmatrix}
	f_1 -h_1 & f_2 -h_2  -if_3 & g_1 & g_2 - ig_3 \\
	f_2 -h_2  + if_3  & f_1 -h_1  & g_2 + ig_3 & g_1 \\
	g_1 & g_2 - ig_3 & f_1 +h_1  & f_2 +h_2 -if_3 \\
	g_2 + ig_3 & g_1 & f_2 +h_2  + if_3 & f_1 +h_1 \\
	\end{pmatrix} \begin{pmatrix}
	\alpha \\
	\overline{\alpha } \\
	\beta \\
	\overline{\beta } \\
	\end{pmatrix},
 \label{eq:eigen}
\end{equation}
where overbars denote complex conjugation. Note that the equation is in the Bogoliubov form with the matrix on the right-hand-side being Hermitian. For succinct expressions of the matrix coefficients, let us introduce two distinct orthogonal decompositions of the film normal $\hat{\bm{z}} = z_A \bm{n}_A^0 + z_{\perp A} \hat{\bm{Z}}\times \bm{n}_A^0 + z_Z \hat{\bm{Z}} = z_B \bm{n}_B^0 + z_{\perp B}\hat{\bm{Z}}\times \bm{n}_B^0 + z_Z \hat{\bm{Z}}$, where $z_A = \bm{n}_A^0 \cdot \hat{\bm{z}} , z_{\perp A} = \left( \hat{\bm{Z}}\times \bm{n}_A^0 \right) \cdot \hat{\bm{z}}, z_Z = \hat{\bm{Z}}\cdot \hat{\bm{z}} $ and similarly for the $B$ layer. The coefficients are then given by
\begin{align}
f_1 =&\ \mu _0 \bm{H}\cdot \frac{\left| \gamma _A \right| \bm{n}_A^0 + \left| \gamma _B \right| \bm{n}_B^0}{2} - \frac{1}{2}\left( \frac{\left| \gamma _A \right|}{d_A M_A}+\frac{\left| \gamma _B \right|}{d_B M_B} \right) \left[ J_1 \bm{n}_A^0 \cdot \bm{n}_B^0 +J_2 \left\{ 3\left( \bm{n}_A^0 \cdot \bm{n}_B^0 \right) ^2 -1\right\} \right] \notag \\
&   + \left| \gamma _A \right| \mu _0 M_A \frac{z_{\perp A}^2 + z_Z^2 -2z_A^2 }{4}   + \left| \gamma _B \right| \mu _0 M_B \frac{z_{\perp B}^2 + z_Z^2 -2 z_B^2}{4}   \\
f_2=& \ \left| \gamma _A \right| \mu _0 M_A \frac{z_{\perp A}^2 - z_Z^2}{4} + \left| \gamma _B \right| \mu _0 M_B \frac{z_{\perp B}^2 - z_Z^2}{4}  + \frac{1}{2}\left( \frac{\left| \gamma _A \right|}{d_A M_A} +\frac{\left| \gamma _B \right|}{d_B M_B}\right) J_2 \left\{ 1 - \left( \bm{n}_A^0 \cdot \bm{n}_B^0 \right) ^2 \right\} , \\
f_3=& \ \frac{ \left| \gamma _A \right| \mu _0 M_A z_{\perp A}  + \left| \gamma _B \right| \mu _0 M_B z_{\perp B}}{2} z_Z, \\
g_1=& \ \mu _0 \bm{H}\cdot \frac{\left| \gamma _A \right| \bm{n}_A^0 - \left| \gamma _B \right| \bm{n}_B^0 }{2} - \frac{1}{2}\left( \frac{\left| \gamma _A \right|}{d_A M_A}-\frac{\left| \gamma _B \right|}{d_B M_B} \right)  \left[ J_1 \bm{n}_A^0 \cdot \bm{n}_B^0 + J_2 \left\{ 3\left( \bm{n}_A^0 \cdot \bm{n}_B^0 \right) ^2 -1 \right\}  \right] \notag \\
& +\left| \gamma _A \right| \mu _0 M_A \frac{z_{\perp A}^2 + z_Z^2  - 2z_A^2}{4}  - \left| \gamma _B \right| \mu _0 M_B \frac{z_{\perp B}^2 + z_Z^2 - 2z_B^2}{4}\\
g_2 =& \ \left| \gamma _A \right| \mu _0 M_A \frac{z_{\perp A}^2 - z_Z^2}{4} - \left| \gamma _B \right| \mu _0 M_B \frac{z_{\perp B}^2 - z_Z^2}{4} + \frac{1}{2}\left( \frac{\left| \gamma _A \right|}{d_A M_A}-\frac{\left| \gamma _B \right|}{d_B M_B} \right) J_2 \left\{ 1-\left( \bm{n}_A^0 \cdot \bm{n}_B^0 \right) ^2 \right\} ,\\
g_3=& \ \mu_0 \frac{ \left| \gamma _A \right| M_A z_{\perp A} -\left| \gamma _B \right| M_B z_{\perp B} }{2} z_Z , \\
h_1 =& \ - \sqrt{\frac{\gamma _A \gamma _B}{d_A M_A d_B M_B}} \left[ \frac{1+\bm{n}_A^0 \cdot \bm{n}_B^0}{2} J_1 + \left\{ 2\left( \bm{n}_A^0 \cdot \bm{n}_B^0 \right) ^2 + \bm{n}_A^0 \cdot \bm{n}_B^0 -1\right\} J_2 \right] , \\
h_2=& \ \sqrt{\frac{\gamma _A \gamma _B}{d_A M_A d_B M_B}} \frac{1-\bm{n}_A^0 \cdot \bm{n}_B^0 }{2}\left\{ J_1 + 2\left( 1+ 2\bm{n}_A^0 \cdot \bm{n}_B^0 \right) J_2 \right\} .
\end{align}
The eigenfrequencies of Eq.~(\ref{eq:eigen}) can be calculated as 
\begin{eqnarray}
\omega ^2 &=& f_1^2 -f_2^2 -f_3^2 + g_1^2 -g_2^2 -g_3^2  + h_1^2 -h_2^2 \\
&&  \pm 2 \sqrt{ \left( f_1 g_1 -f_2 g_2 -f_3 g_3 \right) ^2 + \left( f_1 h_1 -f_2 h_2 \right) ^2 - \left( g_1 h_2 - g_2 h_1 \right) ^2 - g_3^2 \left( h_1^2 - h_2^2 \right)  } . \nonumber
\label{eq:eigenf}
\end{eqnarray}
\end{widetext}

One can observe that the ``couplings" $g_{1,2,3}$ between $\alpha $ and $\beta $ all vanish if the two layers are identical and $\bm{H}$ is in the plane. For identical layers with $\theta \neq 90\degree $, $g_1 =g_2 =0, g_3 \neq 0$ due to $z_{A\perp } =- z_{B\perp }$ and the problem reduces to that of Refs.~\cite{macneill2019gigahertz,sud2020tunable}. The variables $\alpha ,\beta $ represent oscillations that are odd and even under $\mathcal{C}_2 \times \{A\leftrightarrow B\}$, and can be considered generalisations of the acoustic and optical modes in SyAFs, respectively. When $g_{1,2,3}$ become comparable with $f_{1,2,3}, h_{1,2}$, however, $\alpha $ and $\beta $ evenly contribute to the eigenmodes for all values of $H$. This makes it meaningless to talk about hybridisation between odd and even modes, which would require the modes be weakly coupled away from a resonance region and come almost degenerate upon tuning some parameters. Indeed, there is no simple relation between $g_{1,2,3}$ and the spectral gap in general.

\begin{figure*}[ht!]
\centering
\includegraphics[width=15cm]{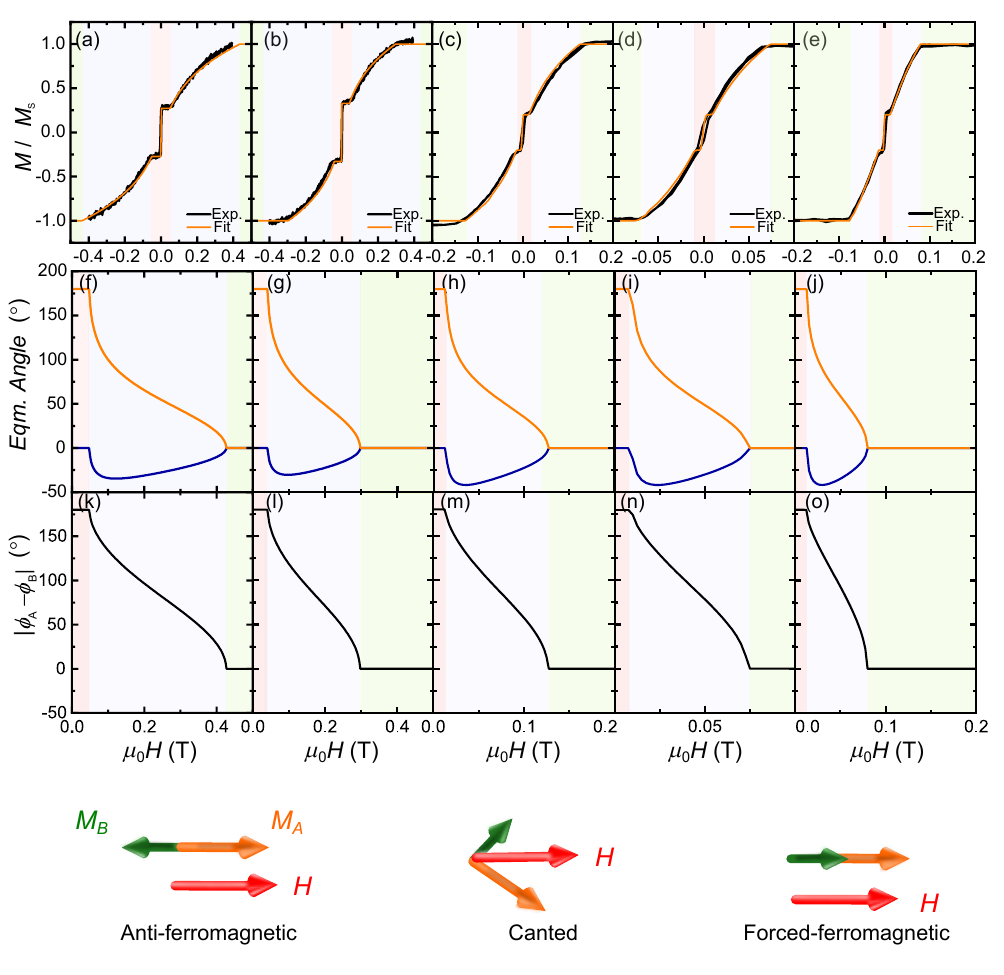}
\caption{(a-e) Normalised $M$-$H$ loops for different set of samples
 (a) NiFe(5)/Ru(0.4)/NiFe(3)
 (b) NiFe(3)/Ru(0.4)/NiFe(5)
 (c) CoFeB(3)/Ru(0.45)/NiFe(3)
 (d) CoFeB(3)/Ru(0.5)/NiFe(3) and 
 (e) CoFeB(3)/Ru(0.55)/NiFe(3). The field is applied along the in-plane easy axis. The solid lines are fit obtained by the theoretical static state calculations based on Eq.~(\ref{eq:free-energy}). (f-j) Static state angles of magnetisation for two FM layers calculated for the best fit parameters corresponding to (a-e) respectively. (k-o) Angle between the two magnetisations. }
\label{fig2}
\end{figure*}
\begin{table*}[ht!]
\centering
\caption{\label{tab:tab1}  Summary of the VSM magnetometry parameters used to obtain the theoretical magnetisation curves shown in Fig.~\ref{fig2} according to Eqs.~(\ref{eq:free-energy}) and (\ref{eq:two}). The left column represents the sample geometry where $".."$ indicates the thermally oxidized $\textrm{Si}$ substrate and the FM near to the substrate is the second FM layer referred to as $B$ layer. $\mu _0 H_{\rm{ex}},\mu _0 H_{\rm 2ex}$ are the quadratic and biquadratic exchange fields respectively and $M_{A(B)}, d_{A(B)}$ are the magnetisation and thickness for the two ferromagnetic layers (NiFe/CoFeB). }
\begin{tabular}{ccccccc}
\hline\hline
Sample& $\mu_{0} M_A$ & $\mu_{0} M_B$ &$ \mu_{0}H_{\rm{ex}}$ &$ \mu_{0} H_{2\rm{ex}}$&$d_A$ &$d_B$\\
geometry &(T) &(T)& (T)&((T)&(nm)&(nm)\\
\hline
 Ta/NiFe/Ru(0.4)/NiFe/Ta/.. &0.95 &0.9 &0.145 &0.03 &5 &3\\
Ta/NiFe/Ru(0.4)/NiFe/Ta/.. &0.95 &0.9 &0.1 &0.02 &3 &5\\
Ta/CoFeB/Ru(0.45)/NiFe/Ta/.. &1.5 &1.0 &0.048 &0.005 &3 &3\\
Ta/CoFeB/Ru(0.5)/NiFe/Ta/.. &1.5 &1.0 &0.02 &0.003 &3 &3\\
Ta/CoFeB/Ru(0.55)/NiFe/Ta/.. &1.5 &1.0 &0.03 &0.002 &3 &3\\
\hline
\hline
\end{tabular}
\end{table*}
\section{Sample growth and magnetometry characterisation} 
Samples used in this study were grown by using magnetron sputtering techniques inside a chamber at a base pressure better than 5$\times$10$\textsuperscript{-6}$ Pa. As summarised in Table~\ref{tab:tab1}, we studied five different multi-layers Ta(5)/FM$_1$($d_\textrm A$)/Ru/FM$_2$($d_\textrm B$)/Ta(5)/thermally oxidized Si~substrate (numbers in the brackets represent
layer thickness in nm) after optimising growth conditions~\cite{Kamimaki_PRAppl2020,sud2020tunable,sud2021parity}. 
Figure~\ref{fig2} shows normalised hysteresis loops for the samples measured for static external field in the plane by vibrating sample magnetometer (VSM) techniques. Three regions distinguished by the alignment of magnetisations of the two layers are indicated in different colours. As explained in the previous section, due to the competition between the exchange and Zeeman energies, our samples undergo two phase transitions. In the small magnetic field limit $H<H_{\rm sf}$ (shadowed in pink), the exchange energy dominates and the two moments are aligned antiferromagnetically. As the field is increased, the spin-flop transition takes place, after which the two moments tilt away from the field in a canted state. Finally, at higher field values $H>H_{\rm ff}$, the Zeeman energy prevails and the magnetic moments point along the field direction entering the forced ferromagnetic regime as indicated in green for each plot. 

Equation~(\ref{eq:free-energy}) was used for fitting to determine the static states of each moment. For obtaining the ground state, we find the values of $\cos\phi_{\rm {A,B}}$ that minimise Eq.~(\ref{eq:free-energy}) for $\theta _A =\theta _B = 90\degree$ in an iterative manner for each magnetic field. 
The orange curves in the first row of Fig.~\ref{fig2} are the normalised magnetisation calculated for each field value as:
\begin{equation}
 \frac{M(H)}{M_s} =\frac{ d_A M_A \cos\phi_A + d_B M_B \cos\phi_B}{ d_A M_A + d_B M_B }.
 \label{eq:two}
\end{equation}
where $M_s $ is the total saturation magnetisation of the sample. Optimisation with respect to the experimental curves yielded the best-fit values of $M_A, M_B$ as well as the quadratic ($ \mu_0H_{\rm{ex}}=J_{1}/\sqrt{d_A d_B M_A M_B}$) and biquadratic exchange fields ( $\mu_0H_{2\rm{ex}}= J_{2}/\sqrt{d_A d_B M_A M_B}$), which are summarised in the Table~\ref{tab:tab1}. While the microscopic origin of $J_1$ is well-explained by the RKKY interaction via electrons in the spacer layer~\cite{STILES_JMMM1999}, the identification of physical origins for $J_2$ is challenging among the several proposals~\cite{demokritov1998biquadratic}, such as intrinsic mechanism~\cite{BARNAS_JMMM1991,Inoue_JMMM1994}, extrinsic fluctuation~\cite{Slonczewski_PRL1991} and magnetic-dipole origin~\cite{Demokritov_PRB1994}. We however mention that our theoretical model and spin dynamics measurements treat the $J_2$ term phenomenologically and are not influenced by its microscopic origin.

\section{Spin dynamics}
\begin{figure*}[hbt!]
\centering
\includegraphics[width=17cm]{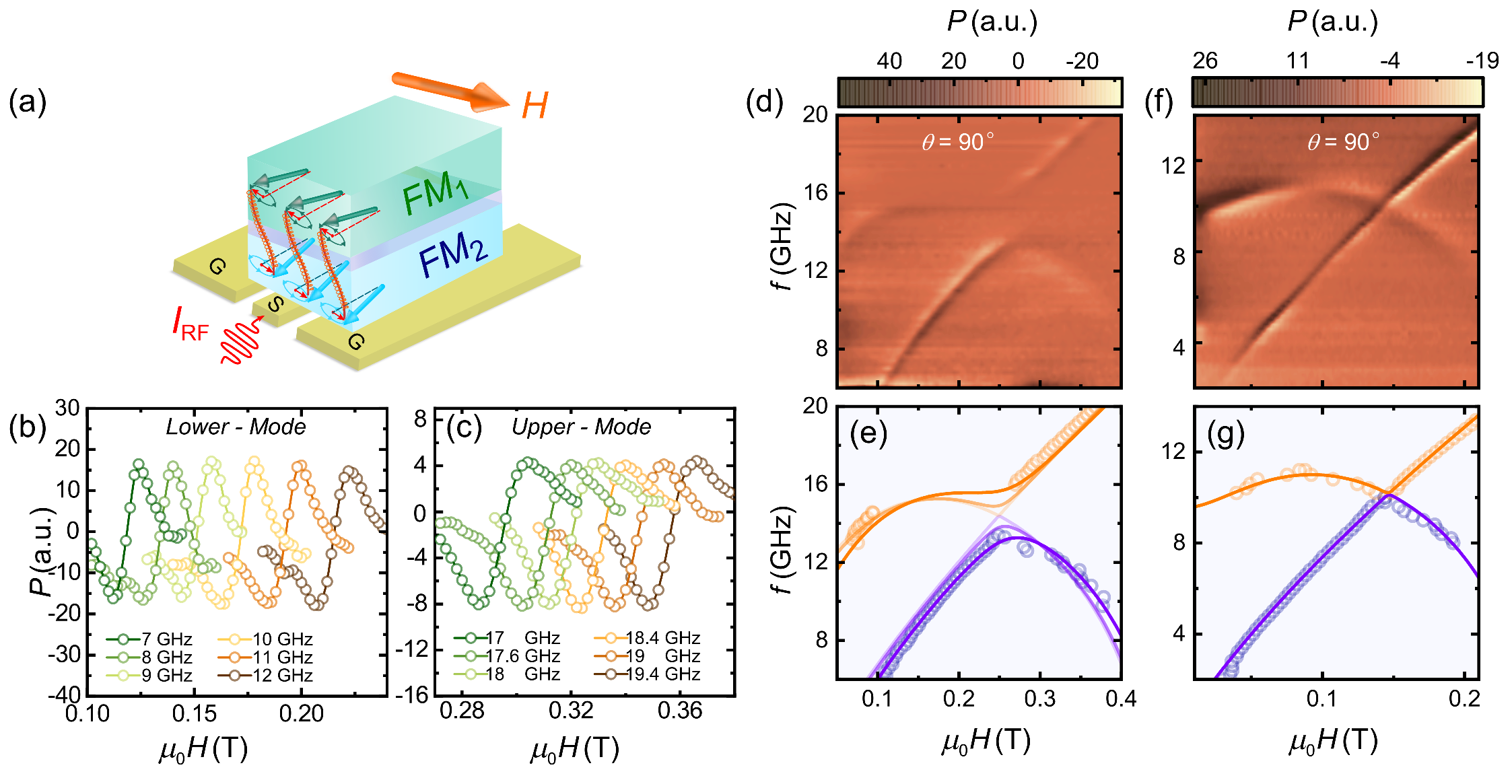}
\caption{(a) Schematic of the sample structure. (b)-(c) Absorption spectra for the sample NiFe(5)/Ru(0.4)/NiFe(3) at (b) low and (c) high field for $\theta$=90$^\circ$. (d) Microwave transmission as a function of frequency and field for the sample NiFe(5)/Ru(0.4)/NiFe(3). The field is applied within the plane, $\theta$=90$^\circ$. A clear avoided-crossing gap is visible at field $\mu _0 H =0.25$ T. (e) Fitting results for data as in (d). The solid lines are fitted curves obtained from macrospin model. The increasing transparencies of the lines correspond to the model calculations for the case ($d_A$,$d_B$)=(5 nm, 3 nm), (5 nm, 4 nm) and (5 nm, 5 nm) respectively. It is seen from the calculations that the spectral gap widens as the thickness asymmetry is increased. (f-g) Similar plots as in (d-e) for sample NiFe(5)/Ru(0.4)/NiFe(5) at $\theta$=90$^\circ$. A clear crossing is seen at at field $\mu _0 H =0.15$ T. This crossing indicates that the two modes are degenerate due to the inter-layer symmetry.
}
\label{fig3}
\end{figure*}
High frequency responses of the coupled moment systems were characterised by broadband on-chip microwave absorption techniques. As illustrated in Fig.~\ref{fig3}(a), each sample chip was placed face-down on a coplanar waveguide~\cite{kalarickal2006ferromagnetic}. For each measurement, we fixed the frequency $f$ and swept a dc external magnetic field $\mu _0 H$ while applying an ac magnetic field at 12 Hz for lock-in detection techniques. Here we show our measurements on the samples NiFe(5)/Ru(0.4)/NiFe(3) and CoFeB(3)/Ru/NiFe(3), both showing avoided crossing~\cite{sud2020tunable,macneill2019gigahertz,shiota2020tunable} due to the asymmetry of thickness and magnetic moment size, respectively.

\subsection{NiFe(5 nm)/Ru(0.4 nm)/NiFe(3 nm)}
\begin{figure*}[hbt!]
\centering
\includegraphics[width=18cm]{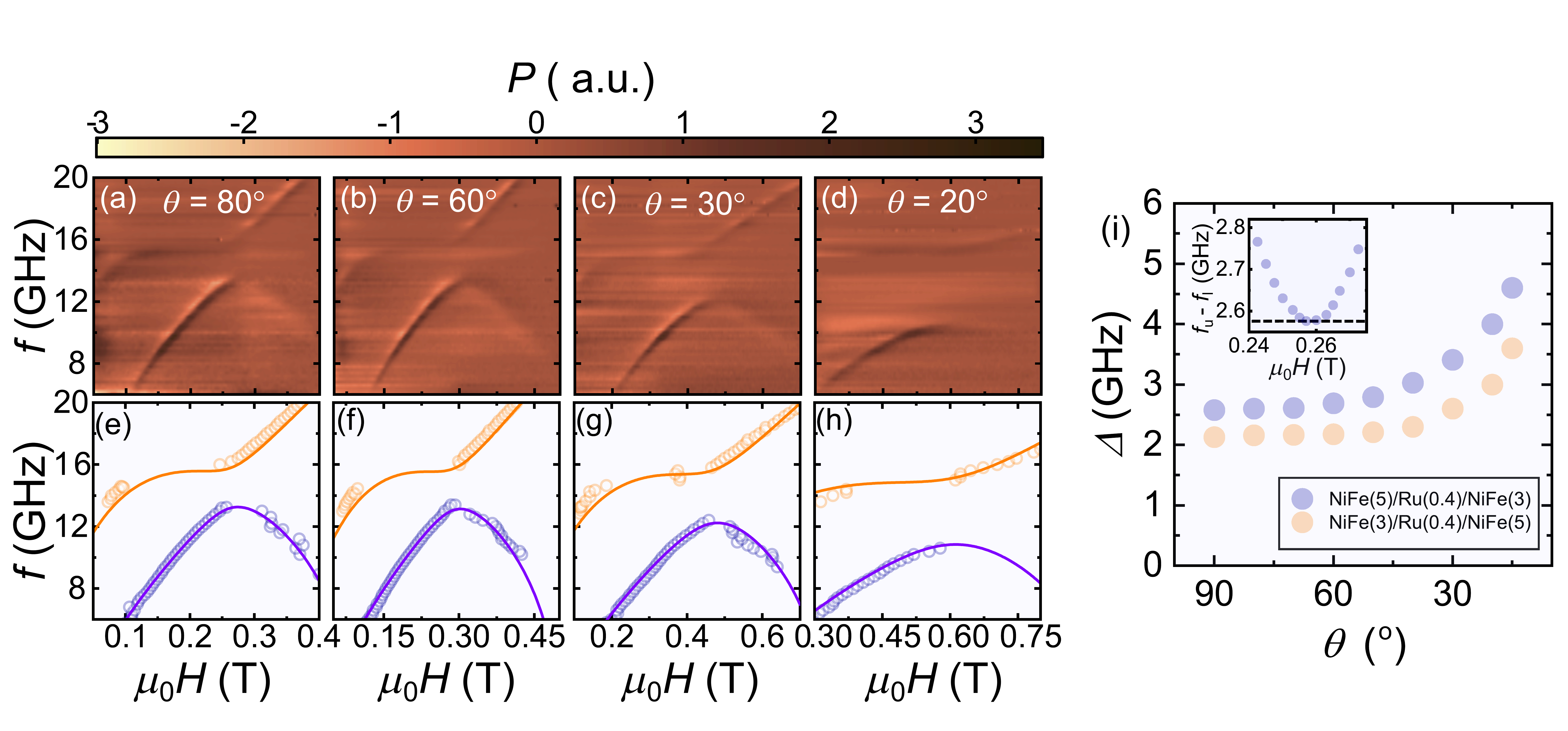}
\caption{(a)-(d) Microwave transmission as a function of frequency and applied field for the sample NiFe(5)/Ru(0.4)/NiFe(3) for different $\theta$. The angle $\theta$ is defined as in Fig.~\ref{fig1}. (e)-(h) Resonance frequency as a function of field obtained by derivative Lorentzian fitting of the experimental data. The solid lines in the figure are theoretical results obtained from the macrospin model. (i) Spectral gap as a function of $\theta$ obtained from theoretical model calculations. It can be seen that a maximum gap of $\approx$ 4.5 GHz is achieved. The spectral gap is defined as the minimum of the difference between the upper ($f_{\textrm{u}}$) and lower ($f_{\textrm{l}}$ ) resonance frequencies as a function of $\mu _0 H$ as shown by the dotted line in inset for the sample NiFe(5)/Ru(0.4)/NiFe(3).  }
\label{fig4}
\end{figure*}

Figures~\ref{fig3}(b) and (c) represent individual measurement curves targeted at two resonance modes in the sample NiFe(5 nm)/Ru(0.4 nm)/NiFe(3 nm) for $\theta = 90^{\circ}$ and different frequencies. These individual scans are used to produce a $f$-$\mu _0 H$ two-dimensional plot as shown in Fig.~\ref{fig3}(d) to capture the absorption spectrum. At $\mu _0 H \approx$ 0.25 T, instead of mode degeneracy, we observe the avoided crossing, suggesting that the in- and out-of-phase oscillations are strongly hybridised~\cite{sud2020tunable, shiota2020tunable}. Figure~\ref{fig3}(e) plots peak positions extracted by individual curve fittings using derivative Lorentzian functions~\cite{rogdakis2019spin,sud2021tailoring,zhao2021growth}. Equation~(\ref{eq:eigenf}) with material parameters independently extracted in the static VSM measurements (Table~\ref{tab:tab1}) generates curves that are in reasonable agreement with experiment. This displays the applicability of the macro-spin model with the minimal set of phenomenological parameters for this type of experiments. To highlight the role of thickness asymmetry for gap opening, we also show two additional sets of model calculations for ($d_A$,$d_B$)=(5 nm, 4 nm) and (5 nm, 5 nm). The model shows that the spectral gap widens as the thickness asymmetry is increased and the gap disappears in a symmetric system. Figures~\ref{fig3}(f-g) confirm this prediction for the symmetric sample NiFe(5 nm)/Ru(0.4 nm)/NiFe(5 nm) with similar thickness of two ferromagnets. The two modes cross each other at $\mu _0 H$ = 0.15 T with an absence of gap in this case. The presence of mode symmetry prevents them from hybridisation and the two modes are degenerate at the crossing point. Due to the asymmetry $d_A \neq d_B $, some of the coupling parameters in the off-diagonal blocks in Eq.~(\ref{eq:eigen}), i.e. $g_1$ and $g_2$, become non-zero, for instance through the prefactor $\left| \gamma _A \right| /\mu _0 M_A - \left| \gamma _B \right| / \mu _0 M_B $). Therefore, even for the case of $\theta = 90\degree$, the thickness asymmetry generates the hybridisation of in- and out-of-phase oscillations. 

We can further increase the gap size by tilting the moments towards the out-of-plane, as we previously demonstrated in the symmetric cases~\cite{sud2020tunable}. Figures~\ref{fig4}(a-d) summarise the experimentally-measured $\theta$ dependence of the magnetic resonances. We performed peak position analysis for these $\theta$-dependent results as shown in Fig.~\ref{fig4}(e-h), together with $\Delta$-$\theta$ relationship plotted in Fig.~\ref{fig4}(i). Here $\Delta $ is defined as the minimum of the difference between the upper and lower resonance frequencies as shown in the Fig.~\ref{fig4}(i) inset. Our theoretical curves successfully reproduce the experimental results without any tunable parameters. As the out-of-plane field increases, the gap is enhanced in comparison with that due to the thickness asymmetry alone and might be attributed to an increase of $g_3$ (Eq.~(16)) with reducing $\theta$. The observed trend is further supported by repeated experiments with a sample with the inverted growth order, i.e. NiFe(3 nm)/Ru(0.4 nm)/NiFe(5 nm), demonstrating approximately the same quantitative behaviour as shown in Figs.~\ref{fig4}(i). This proves that the angle dependence of the gap is a robust feature independent of the assignment of top and bottom layers and small fluctuations in material parameters across different fabrication conditions.

\subsection{CoFeB(3 nm)/Ru/NiFe(3 nm)}
\begin{figure*}[hbt!]
\centering
\includegraphics[width=15cm]{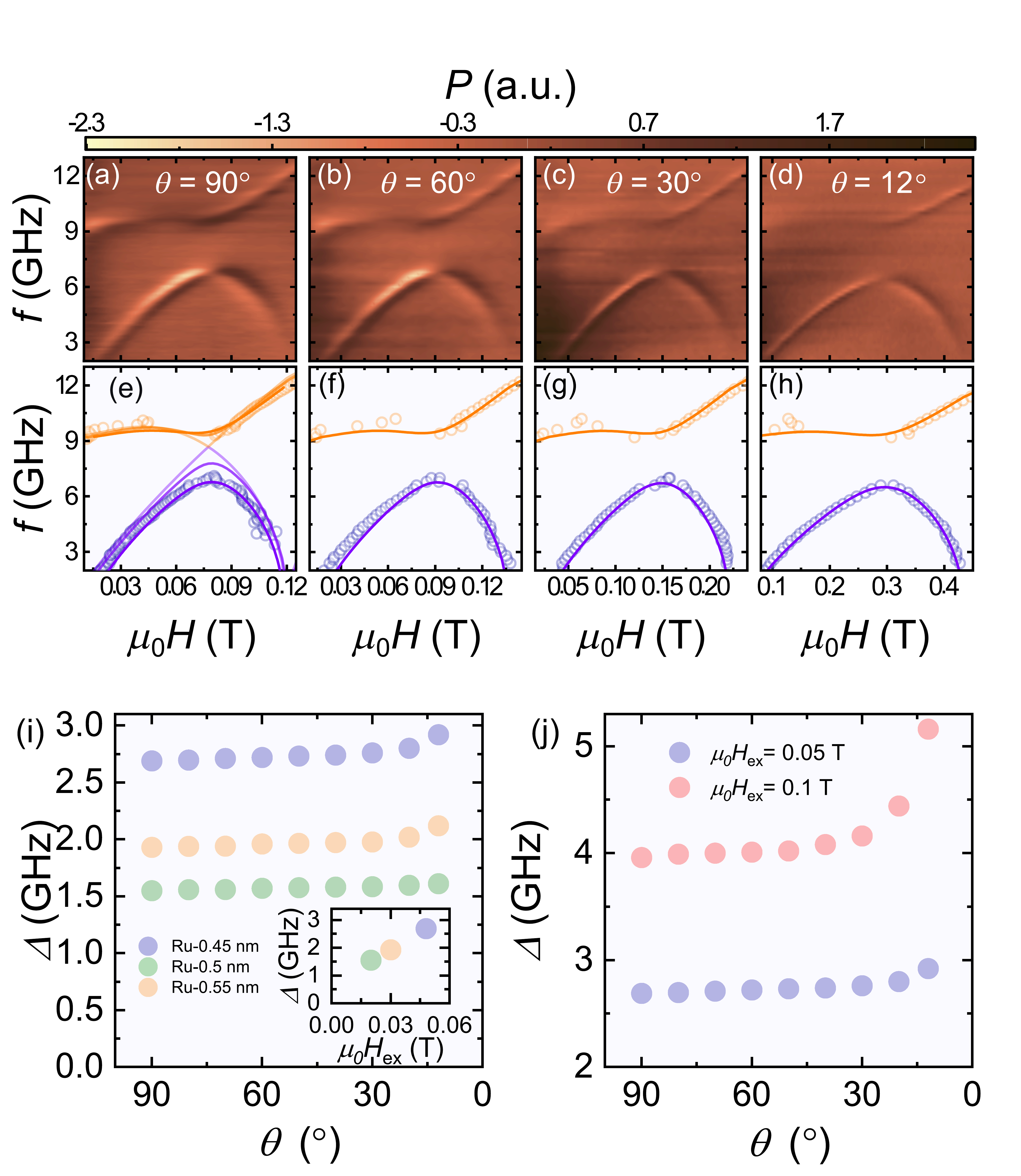}
\caption{(a)-(d) Microwave transmission as a function of frequency and applied field for the sample CoFeB(3)/Ru(0.45)/NiFe(3) for different $\theta$. The spectral gap increases as $\theta$ is decreased. (e)-(h) Resonance frequency as a function of field obtained by derivative Lorentzian fitting of the experimental data. The solid lines in (e)-(h) are theoretical results obtained from the macrospin model. (i) The spectral gap as a function of $\theta$ for different Ru thicknesses, which shows a gradual increase as $\theta$ is decreased. Inset shows the variation of $\Delta$ as a function of $\mu _0 H _\textrm{ex}$. (j) Spectral gap as a function of $\theta$ for sample with Ru thickness 0.45 nm at $\mu _0 H_\textrm{ex}$ = 0.1 T and 0.05 T. The gap shows an increase as $ \mu _0 H_\textrm{ex}$ is increased. The spectra used for extracting the spectral gap is given in Appendix C.
}
\label{fig6}
\end{figure*}
\begin{figure*}[ht!]
\centering
\includegraphics[width=15cm]{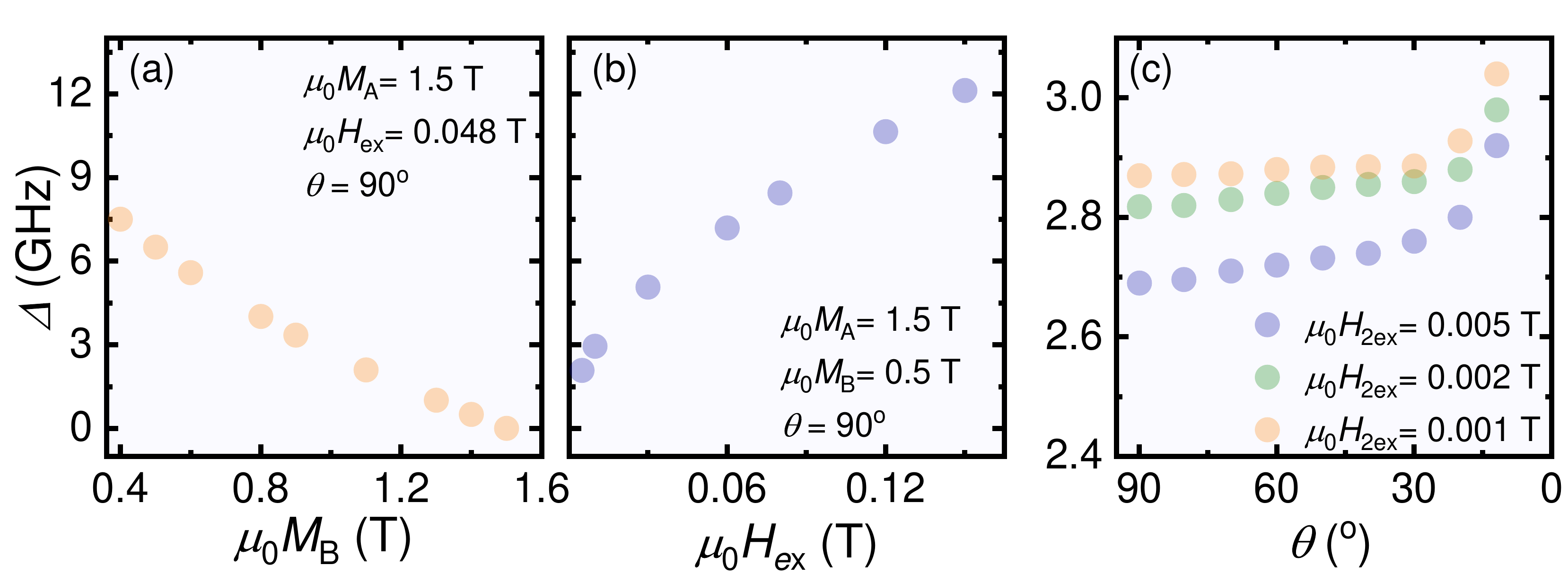}
\caption{Spectral gap obtained from simulations by varying (a) $M_{B}$ and (b) $\mu _0 H_{\textrm{ex}}$. The other fixed parameters used for the simulations are indicated on the plot. As the asymmetry is increased, a very large spectral gap $\approx$ 12 GHz is obtained for $\mu _0 H_{\textrm{ex}}$ = 0.15 T as shown in (b). (c) Spectral gap as a function of $\theta$ for different biquadratic exchange field values $\mu _0 H_{2\textrm{ex}}$ for the sample with Ru thickness of 0.45 nm. The other parameters used for the simulation are the same as given in Table~\ref{tab:tab1}. For low $\mu _0 H_{2\textrm{ex}}$ values, the increase in gap size is not prominent as $\theta $ is varied.}
\label{fig7}
\end{figure*}

In order to experimentally demonstrate the effect of symmetry breaking due to the asymmetry in magnetic moments ($M_{\rm A}\neq M_{\rm B}$)~\cite{PhysRevB.103.064429}, we grew multilayers of CoFeB/Ru/NiFe where the thickness of the two FM materials was kept fixed at 3 nm. Figure~\ref{fig6} summarises the spectral measurements/analysis for the sample CoFeB(3 nm)/Ru(0.45 nm)/NiFe(3 nm) for different values of $\theta$. A clear avoided-crossing gap is visible in the spectra shown in Fig.~\ref{fig6}(a) for $\theta=\pi/2$ and the model calculations (solid curves) reproduce the dispersion curves with the degree of moment asymmetry fixed by the static VSM measurements in this stack as shown in Fig.~\ref{fig6}(e). This is because $g_1$ and $g_2$ become non-zero when $M_A \neq M_B $ (see Eqs.~(14)-(15)). $g_3$ further adds to the coupling when the two moments have out-of-plane components and this tendency is experimentally demonstrated as shown in Figs.~\ref{fig6}(a)-(h). 

Figure~\ref{fig6}(i) displays the gap size $\Delta$ as a function of $\theta$ for the samples CoFeB(3 nm)/Ru($t$)/NiFe(3 nm) with three different Ru thicknesses, i.e. $t=0.45, 0.50$ and 0.55 nm; the magnetic parameters of these samples extracted from VSM measurements are listed in Table~\ref{tab:tab1}. The Ru thickness does not directly enter the free energy equation or LLG equation, instead mostly influencing the interlayer exchange coupling strength $\mu _0 H_\text{ex}$. Hence, comparing these three samples can be a good experimental demonstration of the effect of the exchange coupling strength on GHz spectra for the coupled moments. There is indeed direct correlation between $\Delta$ and $\mu _0 H_\text{ex}$ as shown in the inset of Fig.~\ref{fig6}(i) for $\theta$ = 90$^\circ$.  
We also perform further simulations using the same parameters in the sample CoFeB(3 nm)/Ru(0.45 nm)/NiFe(3 nm), except for $\mu _0 H_\text{ex}$ being 0.1 T. $\Delta$ of this simulation as a function of $\theta$ is plotted in Fig.~\ref{fig6}(j), supporting our claim. 

We have so far shown the reliability of our macrospin model in reproducing the experimental results of magnetic resonance spectra in coupled moments via the interlayer exchange interaction. Here we present our theoretical predictions to discuss the magnetic-parameter dependence of $\Delta$. The asymmetry of coupled moments, i.e. $M_A$ and $M_B$, can be further enhanced in simulation and we find that $\Delta$ is monotonically increased by enlarging the difference between $M_A$ and $M_B$ for a fixed value of $\mu _0 H_{\rm ex}$ as shown in Fig.~\ref{fig7}(a), reaching up to approximately 7.5 GHz with $\mu _0 M_A =1.5$ T and $\mu _0 M_B=0.4$ T. This might be achieved by selecting low-moment magnets as a counterpart of CoFeB to form a stack of synthetic ferrimagnet. Our model simulations also suggest that in such synthetic ferrimagnets with large moment asymmetry, $\mu _0 H_\textrm{ex}$ that can be tuned by the thickness of the intermediate layer can act as a knob to further enhance $\Delta$ as presented in Fig.~\ref{fig7}(b). See Appendix for individual spectra for extracting $\Delta$. Finally, the $\theta$ dependence of $\Delta$ for different values of $\mu _0 H_{2\textrm{ex}}$ is plotted in Fig.~\ref{fig7}(c). For these simulations, an increase of $\mu _0 H_{2\textrm{ex}}$ decreases $\Delta$, which is qualitatively different from the role of $\mu _0 H_{\textrm{ex}}$ on $\Delta$, e.g. in Fig.~\ref{fig6}(j). This is partially because of the general competition between $J_1$ and $J_2$ which prefer different static state configurations and therefore combine to soften the order and decrease the scale of resonance frequency. While $\mu _0 H_{2\textrm{ex}}$ is not a material parameter that can be easily tuned by growth conditions, it is interesting to notice that the biquadratic nature enters the spectral responses very differently from the quadratic counterpart. In general, when the off-diagonal block elements $g_1 ,g_2 ,g_3$ become comparable with the diagonal block ones as in the present case, the notion of coupling between acoustic and optical modes becomes inappropriate, leading to the complex dependence of $\Delta $ on not only the asymmetry related parameters but also the symmetry-respecting ones such as $\mu _0 H_{\rm ex}$ and $\mu _0 H_{2{\rm ex}}$. We would also like to add that in our model, we did not include the mutual spin pumping term between the two magnetic layers~\cite{Chiba_PRB2015}. However, the fact that we have good agreement between experiment and theory without the term indicates that the contribution of the spin-pumping term seems to be insignificant.

\section{Conclusion}
We studied the dynamics of synthetic ferrimagnets and theoretically and experimentally showed their magnon-magnon coupling with dissimilar material and thickness of two ferromagnetic layers. We presented analytical expressions of the coupled mode resonance frequencies and used them to quantitatively discuss experimental results. Using the rich and controllable spin-wave spectra in interlayer-coupled magnetic moments, these materials might find their important use for future magnonic/spintronic applications~\cite{Pirro_NRevMater2021,Chumak_NPhys2015,YUAN_PhysRep2022,ZARERAMESHTI_PhysRep2022,Awschalom_IEEETQM2021,Chumak_IEEE2022}.

\section*{Acknowledgements}
A. S. thanks JSPS Postdoctoral fellowship for research in Japan (P21777) and EPSRC for their supports through NPIF EPSRC Doctoral studentship (EP/R512400/1) during her PhD at UCL. K. Y. is supported by JST PRESTO Grant No.~JPMJPR20LB, Japan and JSPS KAKENHI (No. 21K13886). SM thanks to CSRN in CSIS at Tohoku Univ. and to JSPS KAKENHI (No. 21H04648, 22F21777, 22KF0030)

\appendix

\section{Collinear ground states}
The coordinate axes we used in the main text, Eq.~(\ref{eq:frame}) are not well-defined for $\bm{n}_A^0 \cdot \bm{n}_B^0 =\pm 1$, namely when the two magnetisations are collinear in the static state. This happens for $H\leq H_{\rm sf} $ and $H\geq H_{\rm ff}$ if the magnetic field is in-plane $\theta = 90\degree$, and more generally at high fields if the two layers are identical. 

Let us first discuss the antiferromagnetic state $\bm{n}_A^0 \cdot \bm{n}_B^0 =-1$, for which we can assume $\bm{H}=H\hat{\bm{x}}$. In place of $\mathcal{C}_2$ given in Eq.~(\ref{eq:rotation}), the static state satisfies $\mathcal{C}_2^{\prime }\bm{n}_A^0 = \bm{n}_B^0 $ where
\begin{equation}
    \mathcal{C}_2^{\prime } \bm{v} = \left( \hat{\bm{y}}\cdot \bm{v} \right) \hat{\bm{y}} -\bm{v} .
\end{equation}
One may still then define $\bm{\delta }_{\pm } = \left( \bm{\delta }_A \pm \mathcal{C}_2^{\prime } \bm{\delta }_B \right) /\sqrt{2}$ and decompose them as $\bm{\delta }_{\pm } = \delta _{\pm }^{\perp z} \hat{\bm{z}} \times \bm{n}_A^0 + \delta _{\pm }^{\parallel z} \hat{\bm{z}}$. The rest does not have to be changed with $z_A =z_{\perp A} = z_B = z_{\perp B}=0 , z_Z = 1$ and $\bm{n}_A^0 = -\bm{n}_B^0 = \pm \hat{\bm{x}}$ according to $d_A M_A \gtrless d_B M_B$. 

For the ferromagnetic state $\bm{n}_A^0 \cdot \bm{n}_B^0 =1$, $\hat{\bm{X}} =\bm{n}_A^0$ is well-defined and one may redefine $\hat{\bm{Y}} = \hat{\bm{y}}$. With this provision, $\mathcal{C}_2$ is simply a two-fold rotation about $\hat{\bm{X}}$ and $\bm{\delta }_{\pm } = \bm{\delta }_A \mp \bm{\delta }_B$. Again nothing needs to be modified in Eq.~(\ref{eq:eigen}) and beyond with $z_{\perp A} =z_{\perp B}=0 $.

\section{Additional magnetisation-dynamics results for other samples measured in this study}
\begin{figure}[ht!]
\centering
\includegraphics[width=8.65cm]{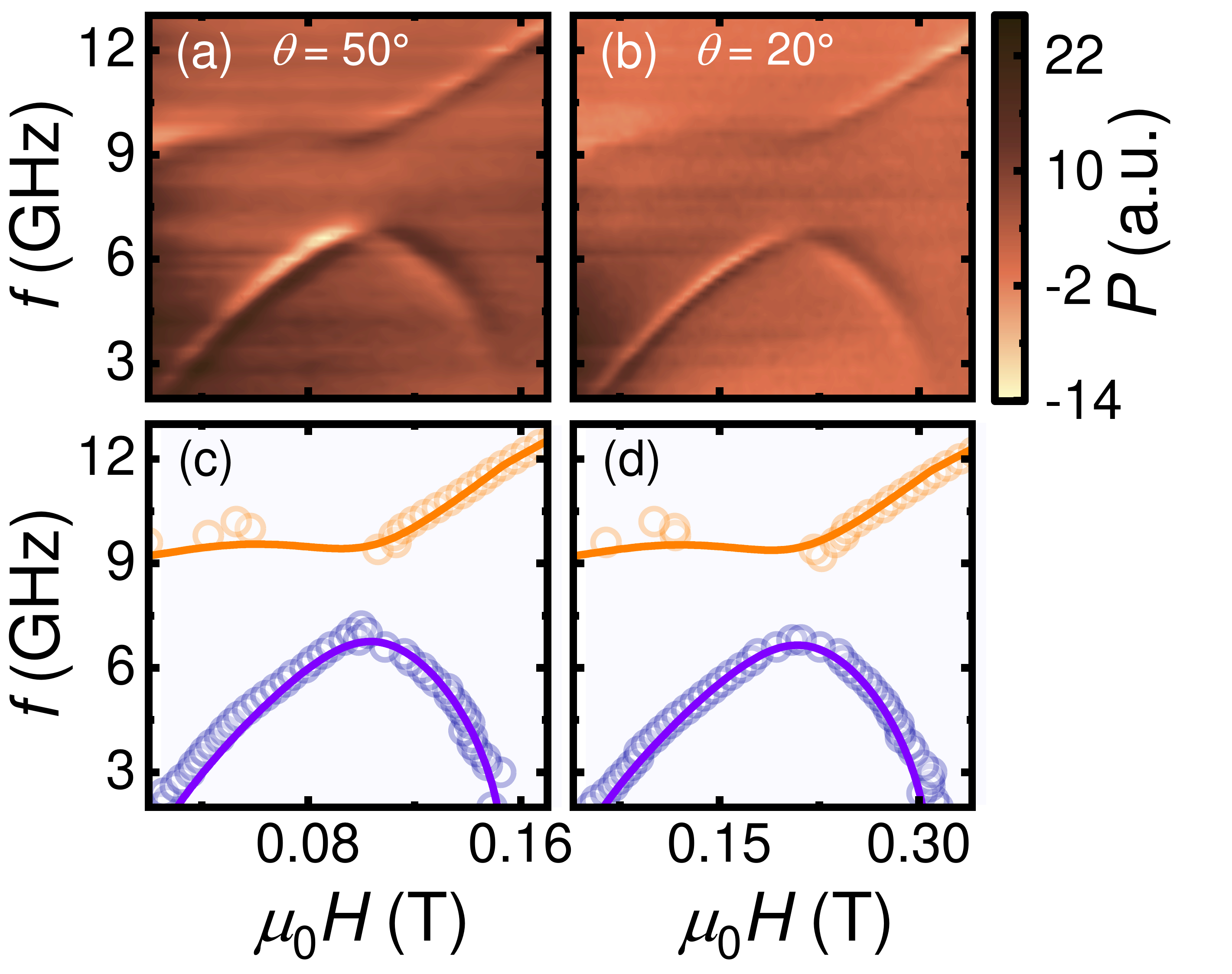}
\caption{(a)-(b) Extra plots of microwave transmission as a function of frequency and applied field for CoFeB(3)/Ru(0.45)/NiFe(3) for different $\theta$ values. The spectral gap increases as $\theta$ is decreased. Figures (c-d) shows resonance frequency obtained using derivative Lorentzian fitting of the experimental data and the solid lines are the theoretical curves obtained from macrospin model.}
\label{fig:S1}
\end{figure}

\begin{figure}[ht!]
\centering
\includegraphics[width=8.65cm]{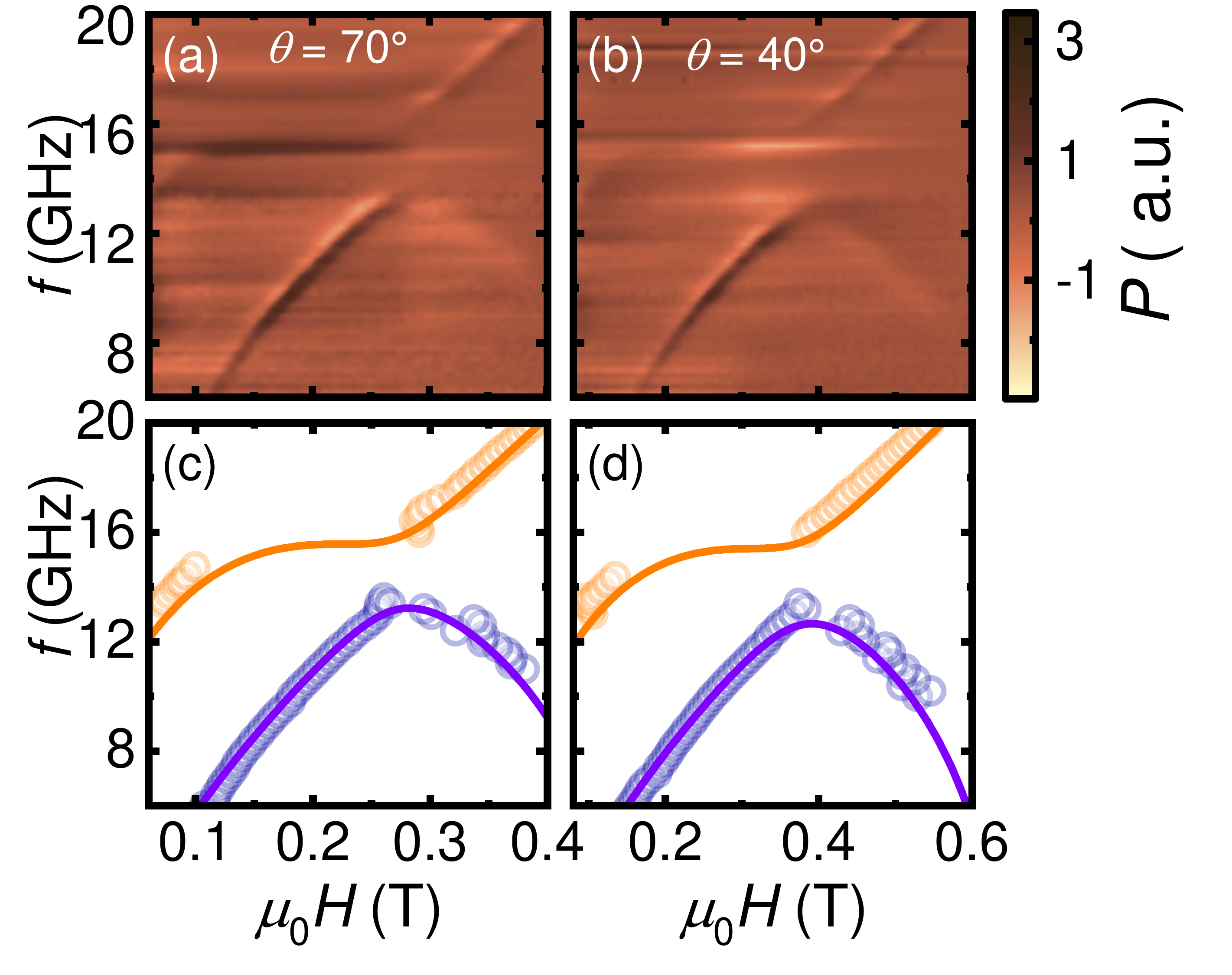}
\caption{(a)-(b) Extra plots of microwave transmission as a function of frequency and applied field for NiFe(5)/Ru(0.4)/NiFe(3) for different $\theta$ values. Figures (c-d) shows resonance frequency obtained using derivative Lorentzian fitting of the experimental data and the solid lines are the theoretical curves obtained from macrospin model for the experimental data as in (a-b).
}
\label{fig:S1(b)}
\end{figure}

\begin{figure}[ht!]
\centering
\includegraphics[width=8.65cm]{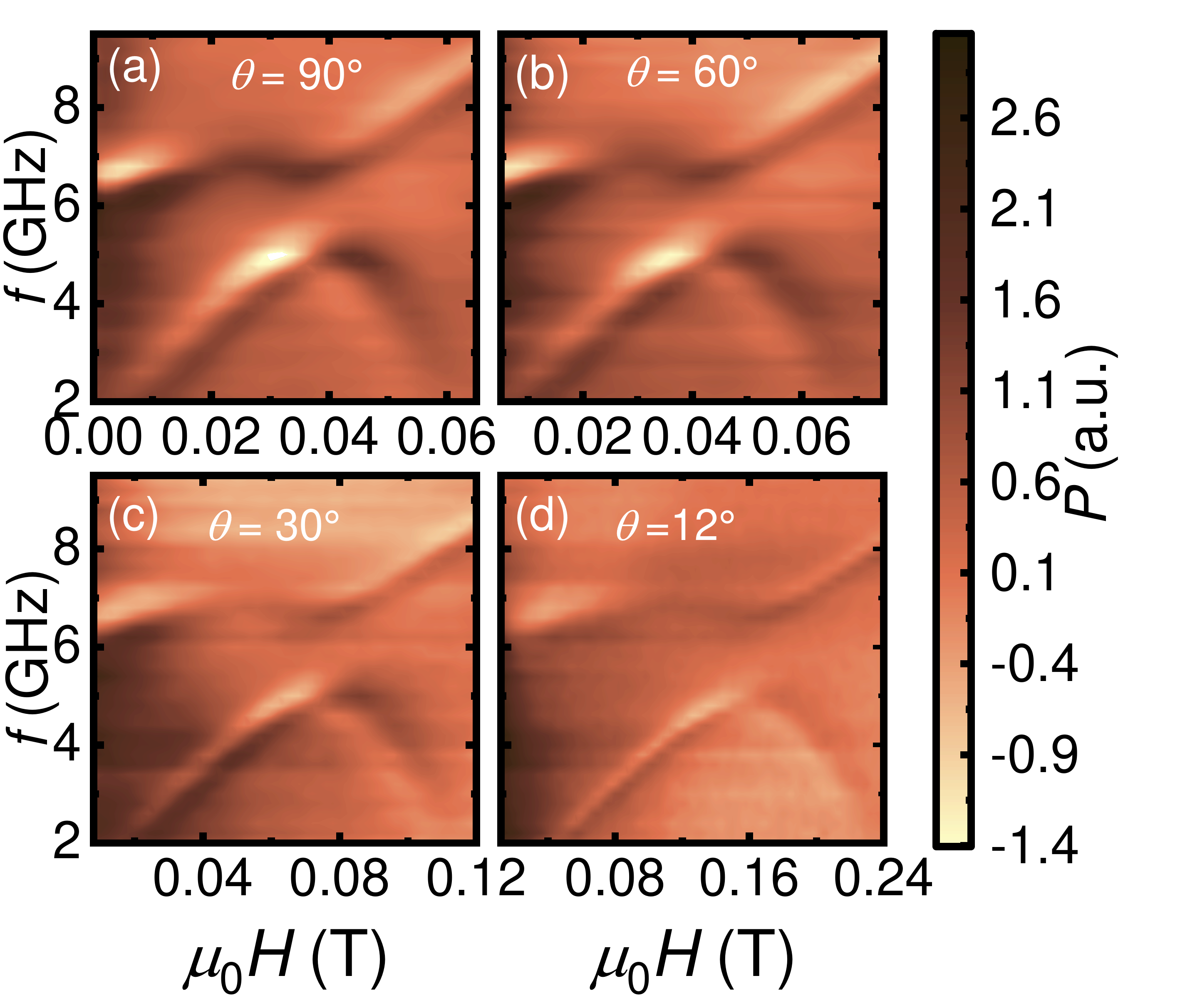}
\caption{(a)-(d) Microwave transmission as a function of frequency and applied field for CoFeB(3)/Ru(0.5)/NiFe(3) for different $\theta$. The gap opening is smaller as compared to sample with Ru thickness 0.45 nm due to smaller $\mu_0 H_{\textrm{ex}}$ of this sample.}
\label{fig:S2}
\end{figure}

\begin{figure}[ht!]
\centering
\includegraphics[width=7.65cm]{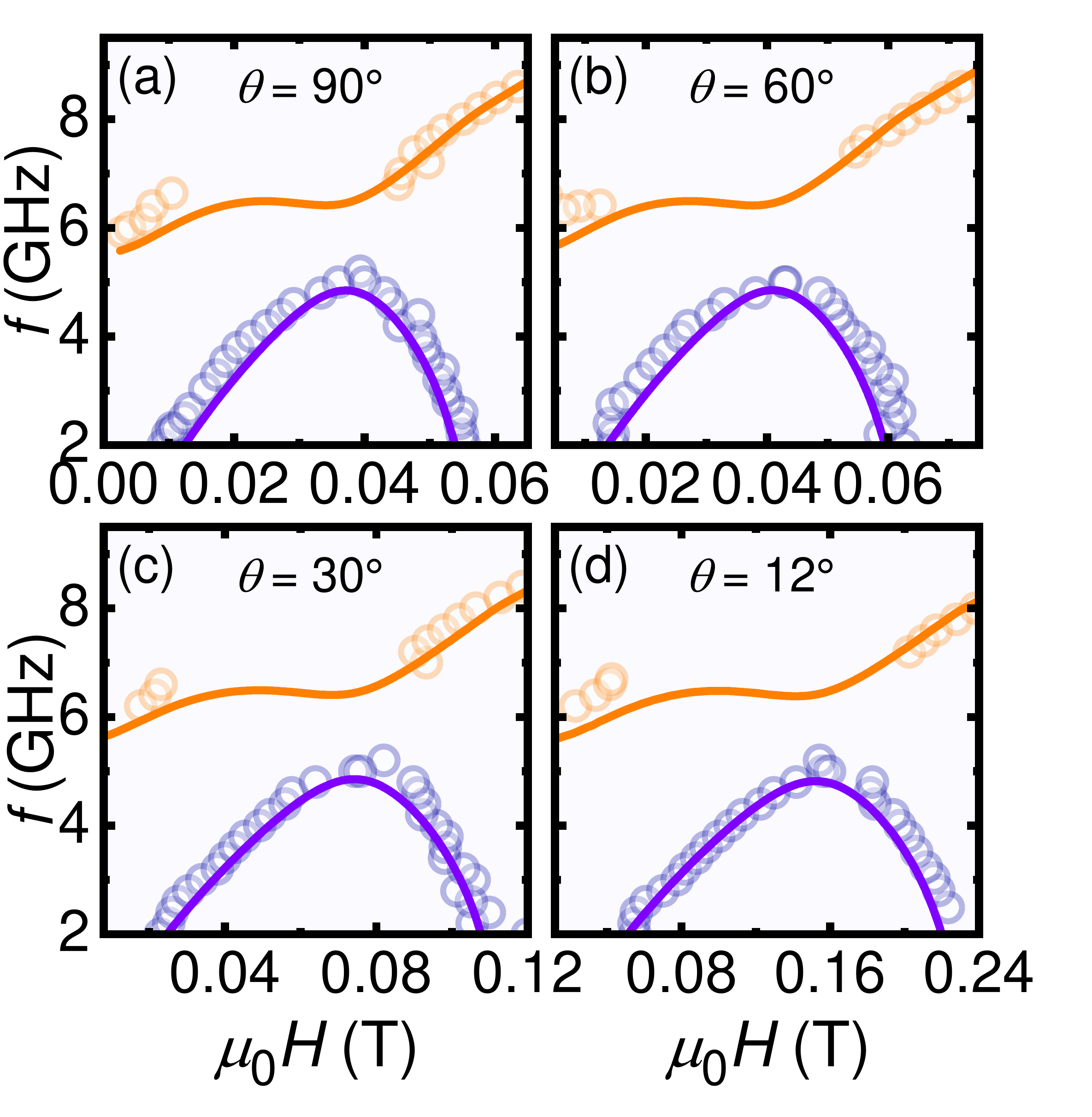}
\caption{(a)-(d) Resonance frequency extracted from derivative Lorentzian fitting of experimental data as a function of applied field along with theoretical prediction for CoFeB(3)/Ru(0.5)/NiFe(3). These correspond to the data shown in Fig.~\ref{fig:S2}.
}
\label{fig:S2(b)}
\end{figure}

\begin{figure}[ht!]
\centering
\includegraphics[width=8.65cm]{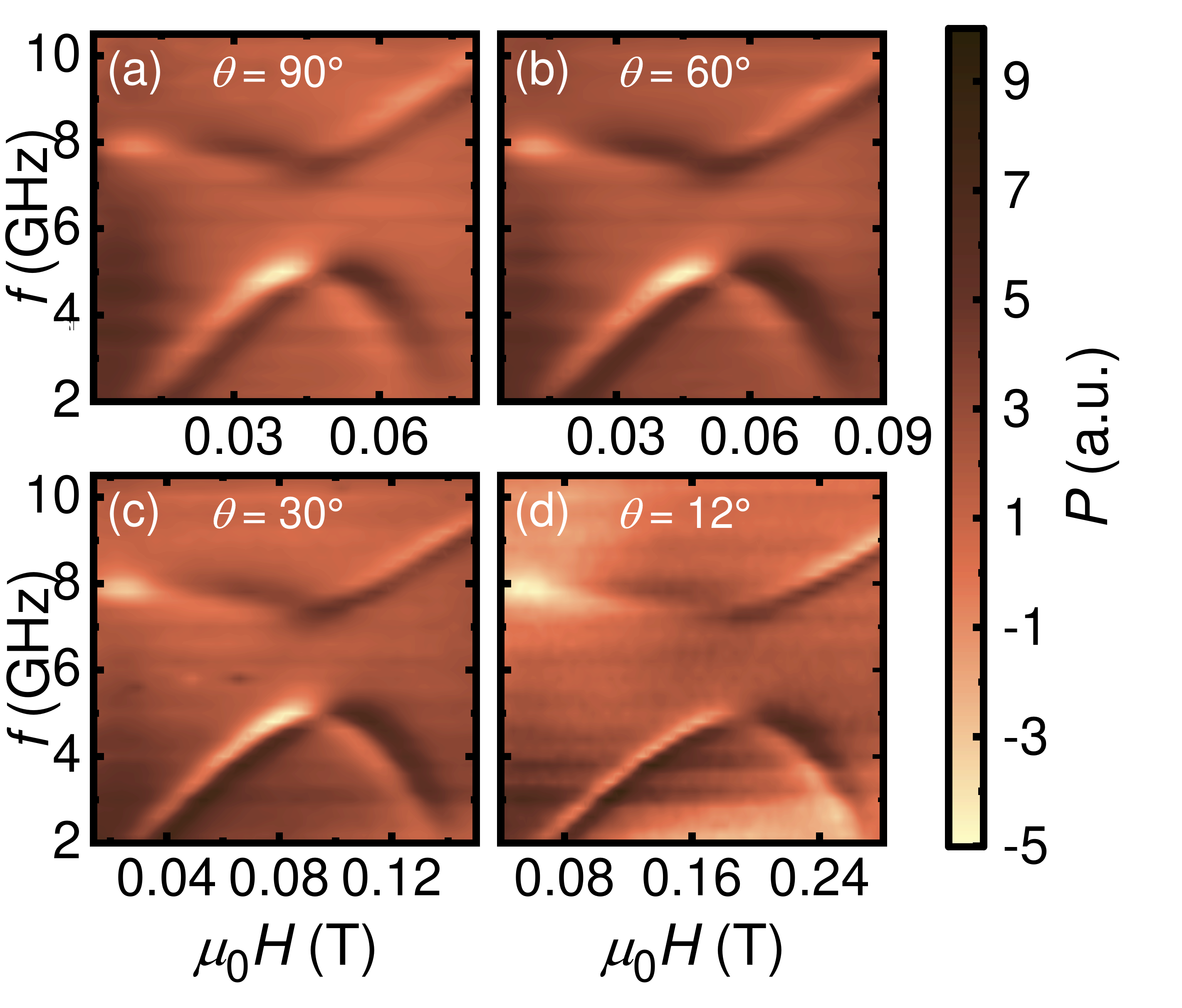}
\caption{(a)-(d) Microwave transmission as a function of frequency and applied field for CoFeB(3)/Ru(0.55)/NiFe(3) for different $\theta$ values. A small variation in  spectral gap is seen as the $\theta$ varied. }
\label{fig:S3}
\end{figure}

\begin{figure}[ht!]
\centering
\includegraphics[width=7.4cm]{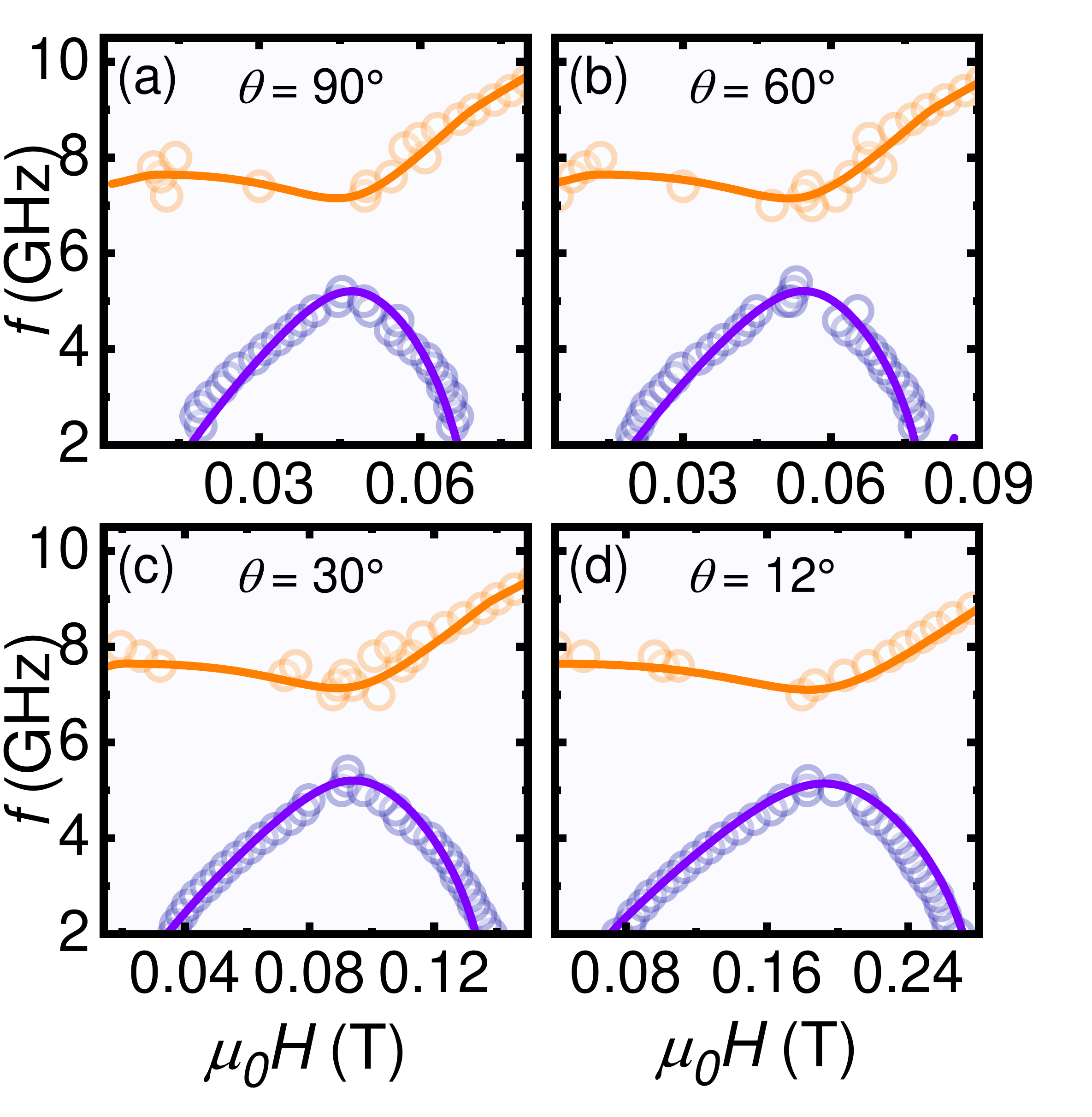}
\caption{(a)-(d) Resonance frequency extracted from fitting of experimental data as a function of applied field along with theoretical prediction for CoFeB(3)/Ru(0.55)/NiFe(3). These correspond to the data shown in Fig.~\ref{fig:S3}.
}
\label{fig:S3(b)}
\end{figure}
This section provides supplementary results for samples measured in this study, which further supports the observations and claims as described in the main text. Top panels (a-b) in Figs.~\ref{fig:S1} and \ref{fig:S1(b)} show measurements for some remaining angles not shown in the main text for the samples with stacking pattern CoFeB/Ru(0.45)/NiFe and NiFe(5)/Ru(0.4)/NiFe(3) respectively. The fittings produced by our macrospin model are shown in bottom panel, which agree well with the experimental data. 

The measurements were repeated for other sets of samples following the procedure outlined in the main text and we saw similar behaviours of spectral gap variation with change in applied field angle towards out-of-plane as shown in Fig.~\ref{fig:S2} for sample CoFeB/Ru(0.5)/NiFe and Fig.~\ref{fig:S3} for sample CoFeB/Ru(0.55)/NiFe. The resonance frequency obtained by fitting of experimental data using derivative of Lorentzian function along with the theoretical predictions are plotted in Figs.~\ref{fig:S2(b)} and \ref{fig:S3(b)} corresponding to Figs.~\ref{fig:S2} and \ref{fig:S3} respectively. For samples with Ru thickness 0.5 and 0.55 nm the gap opening is smaller than the sample with Ru thickness 0.45 nm as shown in Fig.~\ref{fig:S2} and  Fig.~\ref{fig:S3}. This is due to the lower value of $\mu_0 H_{\textrm{ex}}$ in these samples. These results further support our observation of spectral gap controlled by the out-of-plane angle $\theta$ and exchange field strength $\mu_0 H_{\textrm{ex}}$ as mentioned in the main text. 

\section{Numerical simulations to study the impact of varying parameters on coupling gap}

\begin{figure*}[ht!]
\centering
\includegraphics[width=14.4cm]{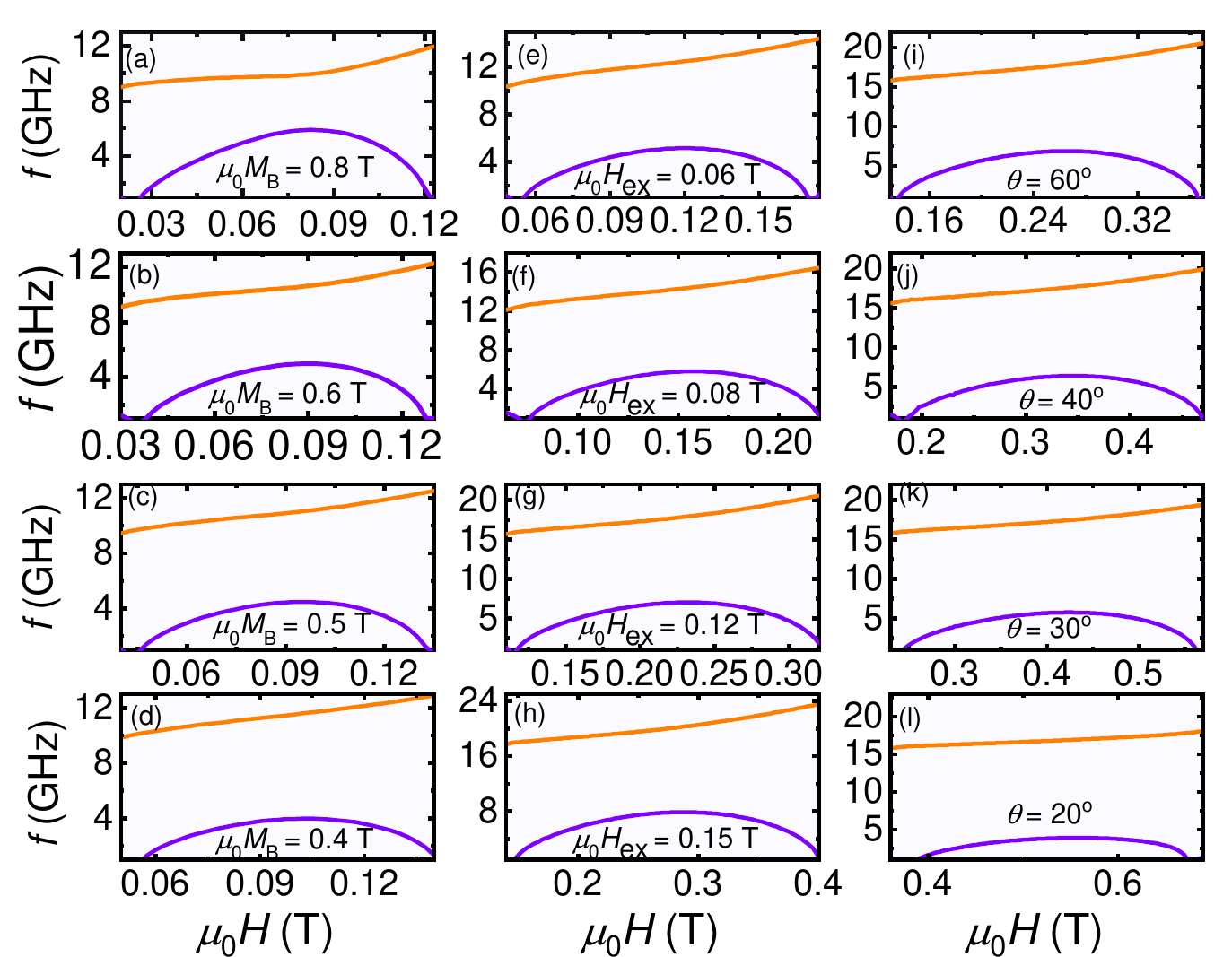}
\caption{Resonance frequency simulation as a function of field obtained by varying different parameters; (a-d) $\mu_0 M_{\textrm{A}}$, (e-h) $\mu_{0} H_\textrm{ex}$, and (i-l) $\theta$. The other parameters which are kept fixed are $\mu_0 M_{\textrm{A}}$ = 1.5 T for all cases. For (a-d) $ \mu_{0} H_\textrm{ex}$ = 0.048 T, $\theta$ = 90$^\circ$, (e-h) $\mu_0 M_{\textrm{B}}$ = 0.5 T, $\theta$ = 90$^\circ$ and (i-j) $\mu_0 M_{\textrm{B}}$ = 0.5 T, $\mu_{0} H_\textrm{ex}$ = 0.12 T. A maximum spectral gap of 12.9 GHz is obtained for (l).
}
\label{figS6}
\end{figure*}

Using numerical simulations, we explored different parameter regimes beyond the experimental conditions. In an effort to understand the magnetic-parameter dependence of $\Delta$ we performed numerical simulations by varying different parameters~$\mu_0 M_{\textrm{A}}$, $\mu_{0} H_\textrm{ex}$ and $\theta$ as shown in Fig.~\ref{figS6}. 

$\Delta$ corresponding to Fig.~\ref{figS6} is shown in Fig.~\ref{fig7} given in the main text. Our numerical simulations suggest that we can tune $\Delta$ by varying different parameters.  

\bibliographystyle{apsrev4-2}
\providecommand{\noopsort}[1]{}\providecommand{\singleletter}[1]{#1}%

\end{document}